\documentclass[aps,a4paper,superscriptaddress,twocolumn,showpacs,preprintnumbers,amsmath,amssymb]{revtex4}
\usepackage{graphicx}
\usepackage{dcolumn}
\usepackage{color}
\usepackage{latexsym,amsfonts}
\usepackage{textcomp}
\usepackage{bm}
\usepackage{hyperref}
\usepackage{ulem}

\def\beq{\begin{equation}}
\def\eeq{\end{equation}}

\begin{document}

\title{Ultra-cold neutrons in qBounce experiments as  laboratory\\ for
  test of chameleon field theories and cosmic acceleration}

\author{Altarawneh,~D.} 
\affiliation{Department of Applied Physics, Tafila Technical University, Tafila 66110, Jordan} 
\email{Correspondence: derar@ttu.edu.jo}

\author{H\"ollwieser,~R.} 
\affiliation{Atominstitut, Technische Universit\"at Wien,
  Stadionallee 2, A-1020 Wien, Austria} 

\date{\today}

\begin{abstract}
The accelerating expansion of the Universe, attributed to dark energy, has spurred interest in theories involving scalar fields such as chameleon field theories. These fields, which couple to matter with density-dependent effective mass, offer a promising explanation for cosmic acceleration. Experiments leveraging ultra-cold neutrons (UCNs) provide an innovative approach to testing these theories.
The existence of a chameleon field, being responsible for the current
phase of cosmic acceleration, is investigated by analysing a free fall
of ultra-cold neutrons from the gap between two mirrors after their
bouncing between these two mirrors. We analyse a deformation of the
wave functions of the quantum gravitational states of ultra-cold
neutrons, induced by a chameleon field, and find a new upper bound
$\beta \le 6.5\times 10^8$ on the chameleon-matter coupling constant
$\beta$ from the unitarity condition. This result refines previous estimates and highlights the potential of ultra-cold neutron experiments as laboratories for exploring scalar field theories and fundamental physics.
\end{abstract}

\maketitle

\section{Introduction}
\label{sec:introduction}

In spite of the confirmation of the existence of the cosmic
acceleration and a cosmological constant $\Lambda$ \cite{Riess} and
the simplest explanation of the cosmological constant
\cite{ECA1, ECA2, ECA3, ECA4}, there remain some questions \cite{ChamDE1},
which may find the answers within the single scalar field theory - the
chameleon field theory \cite{Chameleon1,Waterhouse}. A chameleon field
has been suggested to drive the current phase of cosmic acceleration
for a large class of scalar potentials. A chameleon field coupled to
an ambient matter acquires an effective mass, depending on the matter
density, and becomes heavier in the more densed matter. Thus, the
properties of a chameleon field depend on the density of a matter to
which it is immersed. Because of this sensitivity on the environment,
the scalar field was called {\it chameleon}, and it can couple
directly to baryons and leptons with gravitational strength on Earth
but it would be essentially massless on solar system scales
\cite{Chameleon1,Waterhouse}.

Dark energy, interpreted as negative pressure causing the Universe's acceleration, may originate from a chameleon field, which is responsible for inflating the
Universe~\cite{Silvestri2009,Caldwell2009,Copeland2006,Wetterich1988}. A
competing possibility is that the acceleration is due to a
modification of gravity, i.e., the left-hand side of Einstein's
equation rather than the right. Observations, which can distinguish
these two possibilities are desirable, since measurements of expansion
kinematics alone are not able to.

As has been pointed out in \cite{Brax1,Ivanov2012,Jenke2012}, ultra-cold
neutrons, bouncing in the gravitational field of the Earth above a
mirror \cite{Brax1} and between two mirrors \cite{Ivanov2012,Jenke2012},
can be a good laboratory for testing the existence of a chameleon
field. As has been shown theoretically \cite{qBouncer} and observed
experimentally \cite{qB2002, qB2003, qB2005, qB2007}, ultra-cold neutrons in
the gravitational field of the Earth possess quantum gravitational
bound states with binding energies of order $10^{-12}\,{\rm eV}$. The
transitions between quantum gravitational states of ultra-cold
neutrons, bouncing in the gravitational field of the Earth above a
mirror and between two mirrors, were measured in \cite{AbeleQB1} and
\cite{Jenke2011,Jenke2012}, respectively. There has been found
\cite{Ivanov2012,Jenke2012} that in dependence of the sensitivity of the
experimental data on the transition frequencies of the quantum
gravitational states of ultra-cold neutrons for the chameleon-matter
coupling constant, which is usually called $\beta$, there may be
realised the strong and weak coupling regime. The former is in
agreement with the estimates obtained in
\cite{Adelberger2007,Adelberger2009} from the experimental data on the
gravitational torsion balance.

The problem of applications of a chameleon field to the analysis of
the observable phenomena in the universe and in the terrestrial
laboratories concerns strongly of the definition of the
self--interaction of a chameleon field. Such a self--interaction plays
an important role in the mechanism of the spontaneous creation of the
effective mass of a chameleon field. According to
\cite{Chameleon1,Brax2} and \cite{Chameleon2}, for the analysis of the
properties of a massless chameleon field one may use the potentials of
the self--interaction of a chameleon, which we denote as $V(\phi) \sim
\phi^{-n}$ with positive powers $n$ and $V(\phi) \sim \phi^4$,
respective. 

Below we analyse a chameleon field within the
$\phi^{-n}$--theory with the potential of the self--interaction of a
chameleon field $V(\phi) = \Lambda^4(1 + \Lambda^n/\phi^n)$ with
$\Lambda = 2.4\times 10^{-3}\,{\rm eV}$
\cite{Chameleon1,Brax2,Ivanov2012}.  As has been shown in
\cite{AbeleEP}, in the qBounce experiments the E\"otv\"os constraint
$\eta < 10^{-13}$ \cite{Chameleon1} on the Equivalence Principle (EP)
is fulfilled for $n \ge 2$ only. In turn, from the experimental data
on the transition frequencies of quantum gravitational states
\cite{Jenke2012} one could obtain the estimate on the
chameleon--matter coupling constant $\beta < 5 \times 10^9$. 

In this
paper we investigate a contribution of a chameleon field to the wave
function of ultra-cold neutrons moving above a mirror.  This results
in a qualitative phenomenon, related to the deformation of the wave
function of ultra-cold neutrons, which may be observed experimentally
in the Quantum Bouncing Ball (QBB) experiments
\cite{AbeleWF1}--\cite{AbeleWF3}, i.e. a measurement of the time
evolution of the Schr\"odinger wave function of ultra-cold neutrons,
bouncing above a mirror. 

Ultra-cold neutrons move with a
along the $x$--axis with a velocity $v_0$ between two mirrors (upper
yellow) and (lower red) in the spatial region, which we define as
$\textstyle z^2 \le \frac{d^2}{2}$ \cite{Ivanov2012}. Then, at the
edge, which we localise at $x = 0$ \cite{AbeleWF1}--\cite{AbeleWF3},
ultra-cold neutrons fall down from the height $\textstyle h = D -
\frac{d}{2} + h_1$ above a mirror, localised at $\textstyle z = -
D$. The height $h_1 = -\ell_0 \xi_1 = 13.71\,{\rm \mu m}$ is related
to the ground quantum gravitational state with the binding energy
${\cal E}_1 = m g h_1 = - m g \ell_0 \xi_1 = 0.602\,\xi_1\,{\rm peV} =
1.412\,{\rm peV}$, where $\ell_0 = (2 m^2g)^{-1/3} = 5.87\,{\rm \mu
  m}$ (see Appendix A), is the scale of the quantum gravitational
states of ultra-cold neutrons in the gravitational field of the Earth
\cite{qBouncer} and $\xi_1 = - 2.34497$ (see Appendix A and
\cite{Ivanov2012}). For the numerical analysis we take $v_0 = 6\,{\rm
  m/s}$ \cite{AbeleWF3}, $d = 25.5\,{\rm \mu m}$ \cite{Jenke2011} and
$\textstyle D - \frac{d}{2} = 47\,{\rm \mu m}$. Since the binding
energy of the first excited gravitational state of ultra-cold neutrons
is equal to ${\cal E}_2 = m g h_2 = - m g \ell_0 \xi_2 = 2.614\,{\rm
  peV}$ with $\xi_2 = - 4.34177$ corresponds to the height $h_2 = -
\ell_0 \xi_2 = 25.5\,{\rm \mu m}$, the distance $d = 25.5\,{\rm \mu
  m}$ allows to select between two mirrors ultra-cold neutrons only in
the ground gravitational state.

The paper is organised as follows. In section~\ref{sec:wavefunction}
we construct the wave function of ultra-cold neutrons in the ground
quantum gravitational state, moving between two mirrors in the spatial
region $\textstyle z^2 \le \frac{d^2}{4}$, and define the wave
function of ultra-cold neutrons in the spatial region $z \ge -D$ above
a mirror. We show that the main contributions come from the excited
states with a principal quantum number $k = 6$, $k = 7$ and $k = 8$.
In section~\ref{sec:nphi} we calculate a first order correction to the
wave functions of ultra-cold neutrons, induced by a chameleon
field. From a unitarity condition, we obtain a new estimate on the
chameleon--matter coupling constant $\beta \le 6.5 \times 10^8$. We discuss the obtained results in \ref{sec:discussion} and conclude in \ref{sec:conclusion}. In Appendix A we
define the wave function of the ground quantum gravitational state of
ultra-cold neutrons between two mirrors in the spatial region
$\textstyle z^2 \le \frac{d^2}{4}$ and above a mirror in the spatial
region $z \ge -D$, respectively.

\section{Wave function of ultra-cold neutrons above
 a mirror}
\label{sec:wavefunction}

For the analysis of the evolution of the wave function of ultra-cold
neutrons in the spatial region above a mirror we assume that 1)
ultra-cold neutrons start to fall at $t = 0$ and 2) before the fall
the time evolution and the $x$--degree of freedom of ultra-cold
neutrons are described by the Gaussian wave packet with a width
$\delta = 1\,{\rm \mu m}$ \cite{AbeleWF1}--\cite{AbeleWF3}. At time
$t$ the wave function of ultra-cold neutrons takes the form
\begin{eqnarray}\label{eq:1}
\hspace{-0.3in}&&\psi_1(z, x, t) = \psi_1(z)\,(4\pi
\delta^2)^{1/4}\int^{+\infty}_{-\infty}\frac{dp}{2\pi}\,e^{\textstyle
  \,-\frac{\delta^2 (p - p_0)^2}{2}}\nonumber\\
\hspace{-0.3in}&&\times\,e^{\textstyle\,- i ({\cal E}_1 +
  \frac{p^2}{2m})\,t + ipx} = \psi_1(z)\,\Big(\frac{\delta^2}{\pi
  \delta^4_t}\Big)^{1/4}\,e^{\,\textstyle -\frac{(x - v_0t)^2}{2
    \delta^2_t}}\nonumber\\
\hspace{-0.3in}&&\times\,e^{\textstyle\,- i ({\cal E}_1 + \frac{p^2_0}{2m})\,t
  + ip_0x},
\end{eqnarray}
where the wave function $\psi_1(z)$ is given in Appendix A and
$\delta_t = \delta \sqrt{1 + it/m\delta^2}$. The squared absolute
value $|\psi_1(z, x, t)|^2$ is
\begin{eqnarray}\label{eq:2}
\hspace{-0.3in}&&|\psi_1(z, x, t)|^2 =
|\psi_1(z)|^2\,\frac{1}{\sqrt{\pi \Delta^2_t}}\,e^{\,\textstyle
  -\frac{(x - v_0t)^2}{2 \Delta^2_t}},
\end{eqnarray}
where $\Delta_t = \delta \sqrt{1 + t^2/m^2 \delta^4}$.

Since in the spatial region between two mirrors $\textstyle z^2 \le
\frac{d^2}{4}$ ultra-cold neutrons are only in the ground quantum
gravitational state with the wave function and the excited quantum
gravitational states do not exist, the first order correction to the
wave function of the ground quantum gravitational state, induced by a
chameleon field, appears only as a phase shift of the wave function
\cite{Davydov65}.

In the spatial region $z \ge - D$ the pure ground
gravitational state $\psi_1(z,x,t)$ should be treated as a mixed
quantum gravitational state with the wave function of the $z$--degree
of freedom given by
\begin{eqnarray}\label{eq:3}
\hspace{-0.3in}\psi_1(z,t) = \sum_k C_k\Psi_k(z,t),
\end{eqnarray}
where $\psi_1(z,t) = \psi_1(z)\, e^{\, - i{\cal E}_1t}$, $\Psi_k(z,t)
= \Psi_k(z)\,e^{\, - iE_kt}$ with the wave functions $\Psi_k(z)$,
given in Appendix A, and the coefficients $C_k$ are given by
\begin{eqnarray}\label{eq:4}
\hspace{-0.3in}C_k = \int^{+d/2}_{- d/2}dz\,\Psi^*_k(z)\psi_1(z).
\end{eqnarray}
The integrals are calculated over the spatial region $\textstyle z^2
\le \frac{d^2}{4}$, since the wave function $\psi_1(z)$ vanishes in
the spatial region $\textstyle z^2 \ge \frac{d^2}{4}$. The binding
energies of the first quantum gravitational states of ultra-cold
neutrons and the values of the coefficients $C_k$ for $k = 1,2,\ldots,
15$ are adduced in Table I in Appendix A. One may see that the main
contribution to the mixed state comes from the quantum gravitational
states with $k = 6$, $k = 7$ and $k = 8$ with coefficients $C_6 =
0.42$, $C_7 = 0.66$ and $C_8 = 0.55$ and probabilities $P_6 = |C_6|^2
= 0.18$, $P_7 = |C_7|^2 = 0.44$ and $P_8 = |C_8|^2 = 0.30$,
respectively. For the analysis of the first order corrections to the
wave functions of ultra-cold neutrons, caused by a chameleon field, we will use the following approximation of the wave function $\psi_1(z,t)$
\begin{eqnarray}\label{eq:5}
\hspace{-0.3in}\psi_1(z,t) = C_6\Psi_6(z,t) + C_7\Psi_7(z,t) + C_8\Psi_8(z,t),
\end{eqnarray}
which describes the exact wave function with an accuracy of about
$8\,\%$.

\section{Chameleon field correction in the $\phi^{-n}$--theory}
\label{sec:nphi}

The potential of a chameleon field, coupled to ultra-cold neutrons,
takes the form \cite{Brax1,Ivanov2012}
\begin{eqnarray}\label{eq:6}
\hspace{-0.3in}\Phi(z) = \beta\,\frac{m}{M_{\rm Pl}}\,\phi(z),
\end{eqnarray}
where $M_{\rm Pl} = 2.435\times 10^{27}\,{\rm eV}$ is the Planck mass,
$\beta$ is a chameleon-matter coupling constant and $\phi(z)$ is the
profile of a chameleon field in the spatial region $z \ge -D$. As has
been shown in \cite{Jenke2012} the chameleon-matter coupling constant
obeys the constraint $\beta < 5\times 10^9$. As has been shown in
\cite{Brax1,Ivanov2012}, in the spatial region $z \ge - D$ the profile
of a chameleon field, obtained as a solution of the non--linear
equations of motion in the $\phi^{-n}$--theory and in the strong
coupling limit $\beta \ge 10^5$, takes the form
\begin{eqnarray}\label{eq:7}
\hspace{-0.3in}\phi(z) = \Lambda \Big(\frac{n + 2}{\sqrt{2}}\,\Lambda
D\Big)^{\textstyle \frac{2}{n + 2}}\Big(1 + \frac{z}{D}\Big)^{\textstyle \frac{2}{n + 2}},
\end{eqnarray}
where $\Lambda = 2.4\times 10^{-3}\,{\rm eV}$
\cite{Chameleon1,Brax1,Ivanov2012}. For $D = 59.75\,{\rm \mu m}$ we
may define the chameleon field potential as
\begin{eqnarray}\label{eq:8}
\hspace{-0.3in}\Phi(z) = m g \ell_0 C_{\phi}\frac{\phi(z)}{\Lambda},
\end{eqnarray}
where $C_{\phi} = \beta m \Lambda/M_{\rm pl} mg\ell_0 = 1.54\times 10^{-9}\,\beta$ and
\begin{eqnarray}\label{eq:9}
\hspace{-0.3in}\frac{\phi(z)}{\Lambda} = \Big(\frac{n +
  2}{2}\Big)^{\textstyle \frac{2}{n + 2}}\Big(1 +
\frac{z}{D}\Big)^{\textstyle \frac{2}{n + 2}}.
\end{eqnarray}
For the estimate $\beta < 5\times 10^9$ \cite{Jenke2012} the coupling
constant $C_{\phi}$ is $C_{\phi} < 7.7$. The power $n$ of the
potential of a self--interaction of a chameleon field is a free
parameter of the $\phi^{-n}$--theory of a chameleon field. As has been
shown in \cite{AbeleEP}, in the $\phi^{-n}$--theory of a chameleon
field with the upper bound on the chameleon--matter coupling constant
$\beta < 5 \times 10^9$ the E\"otv\"os constraint $\eta < 10^{-13}$
\cite{Chameleon1} is fulfilled in the qBounce experiments for the
powers $n \ge 2$ of the potential of a self--interaction of a
chameleon field. For $n \ge 2$ the accuracy of the approximation
Eq.(\ref{eq:8}) is better than $1.5\,\%$.

\begin{figure}
\centering
\includegraphics[width=\linewidth]{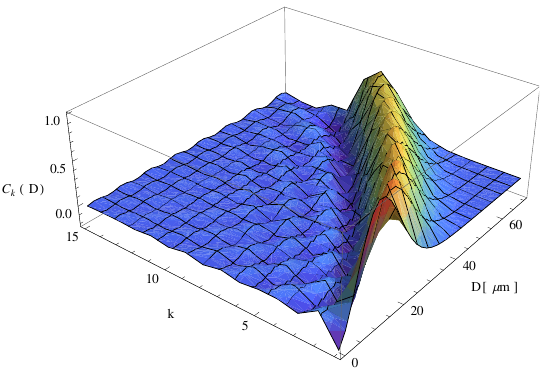}
\caption{Coefficients $C_k(D)$ illustrating their dependence on the distance $D$ and quantum state index $k$.}\label{fig1}
\end{figure}

The wave function $\Psi_k(z,t)$ of ultra-cold neutrons, including the
first order corrections induced by a chameleon field, takes the form
\begin{eqnarray}\label{eq:10}
\hspace{-0.3in}&&\Psi_k(z,t) \to \Psi_k(z,t) + C_{\phi}\sum_{k' \neq
  k}a_{k'k}(n)\,\Psi_{k'}(z,t),
\end{eqnarray}
where we have denoted
\begin{eqnarray}\label{eq:11}
\hspace{-0.3in}a_{k'k}(n) &=& \Big(\frac{\ell_0}{ D}\frac{n +
  2}{2}\Big)^{\textstyle \frac{2}{n + 2}}\frac{\langle
  k'|\xi^{\textstyle \frac{2}{n + 2}}|k\rangle}{\zeta_{k'} -
  \zeta_k},\nonumber\\
\hspace{-0.3in}\langle k'|\xi^{\textstyle \frac{2}{n + 2}}|k\rangle &=& \int^{+\infty}_0
d\xi\, \Psi_{k'}(\xi)\,\xi^{\textstyle \frac{2}{n + 2}}\Psi_k(\xi)
\end{eqnarray}
with $\xi = (D + z)/\ell_0$. For the excited states $\Psi_k(z,t)$ with
the principal quantum number $k = 6$, $7$ and $8$ we obtain the
following corrections
\begin{eqnarray}\label{eq:12}
\hspace{-0.3in}&&\Psi_6(z,t) + C_{\phi}a_{76}(n)
\Psi_7(z,t) + C_{\phi}a_{86}(n) \Psi_8(z,t),\nonumber\\
\hspace{-0.3in}&& \Psi_7(z,t) + C_{\phi}a_{67}(n)
\Psi_6(z,t) + C_{\phi}a_{87}(n)\Psi_8(z,t),\nonumber\\
\hspace{-0.3in}&& \Psi_8(z,t) + C_{\phi}a_{68}(n)
\Psi_6(z,t) + C_{\phi}a_{78}(n)\Psi_7(z,t).
\end{eqnarray}
The wave function  of the mixed state take the form
\begin{eqnarray}\label{eq:13}
\hspace{-0.3in}\psi_1(z,t) \to  \bar{C}_6\Psi_6(z,t) + \bar{C}_7\Psi_7(z,t) + \bar{C}_8\Psi_8(z,t),
\end{eqnarray}
where the coefficients $\bar{C}_k$ for $k = 6$, $7$ and $8$ are equal to
\begin{eqnarray}\label{eq:14}
\hspace{-0.3in}\bar{C}_6 &=& C_6 + C_{\phi}(C_7a_{67}(n) + C_8
a_{68}(n)),\nonumber\\
\hspace{-0.3in}\bar{C}_7 &=& C_7 + C_{\phi}(C_6a_{76}(n) + C_8
a_{78}(n)),\nonumber\\
\hspace{-0.3in}\bar{C}_8 &=& C_8 + C_{\phi}(C_6a_{86}(n) + C_7
a_{87}(n)).
\end{eqnarray}
The coefficients $\bar{C}_k$ must obey the unitarity conditions
\begin{eqnarray}\label{eq:15}
\hspace{-0.3in}&&|\bar{C}_k| < 1\quad {\rm for}\quad k = 6,7,8;\nonumber\\
\hspace{-0.3in}&&|\bar{C}_6|^2 + |\bar{C}_7|^2 + |\bar{C}_8|^2 < 1.
\end{eqnarray}
Such a constraint may be used for the estimate of the chameleon--matter
coupling constant $\beta$. Fig.~\ref{fig1} illustrates the dependence of the coefficients $C_k(D)$ on the distance $D$ and quantum state index $k$.

\begin{figure}
\centering
\includegraphics[width=\linewidth]{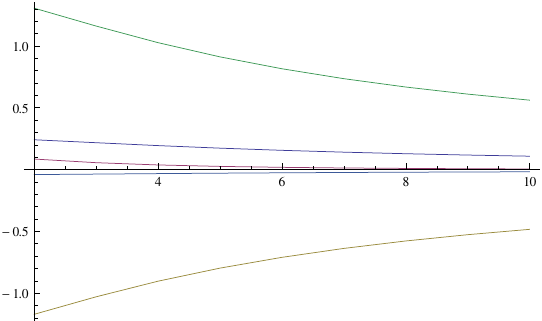}
\caption{The behaviour of $a_{67}(n)$ and $a_{87}(n)$ as functions of $n$ for $n\in [2,10]$.}\label{fig2}
\end{figure}

A numerical analysis shows that the contribution of $a_{86}(n)$ can be
neglected with the contributions of $a_{67}(n)$ and $a_{86}(n)$. Then,
$a_{67}(n) = - a_{76}(n)$ and $a_{87}(n) = - a_{78}(n)$. The behaviour
of $a_{67}(n)$ and $a_{87}(n)$ as functions of $n$ for $n\in [2,10]$
is shown in Fig.~\ref{fig2}. One may see that $a_{67}(n) \simeq -
a_{87}(n)$. Using the obtained relations the coefficients $\bar{C}_k$ are
\begin{eqnarray}\label{eq:16}
\hspace{-0.3in}\bar{C}_6 &=& C_6 + C_{\phi}C_7a_{67}(n),\nonumber\\
\hspace{-0.3in}\bar{C}_7 &=& C_7 + C_{\phi}(C_8 - C_6)a_{67}(n),\nonumber\\
\hspace{-0.3in}\bar{C}_8 &=& C_8 - C_{\phi}C_7 a_{67}(n)
\end{eqnarray}
or
\begin{eqnarray}\label{eq:17}
\hspace{-0.3in}\bar{C}_6 &=& C_6 + 1.54\times
10^{-9}\,\beta\,C_7a_{67}(n),\nonumber\\
\hspace{-0.3in}\bar{C}_7 &=& C_7 + 1.54\times 10^{-9}\,\beta\, (C_8 -
C_6)a_{67}(n),\nonumber\\
\hspace{-0.3in}\bar{C}_8 &=& C_8 - 1.54\times 10^{-9}\,\beta\,C_7
a_{67}(n)
\end{eqnarray}
One may see that to first order perturbation theory with respect to
the coupling constant $C_{\phi} = 1.54\times 10^{-9}\,\beta \ll 1$ the
constraint
\begin{eqnarray}\label{eq:18}
\hspace{-0.3in} |\bar{C}_6|^2 + |\bar{C}_7|^2 + |\bar{C}_8|^2 =
|C_6|^2 + |C_7|^2 + |C_8|^2 < 1
\end{eqnarray}
is fulfilled. However, for the upper bound $\beta < 5\times 10^9$ on
the chameleon--matter coupling constant we obtain that $C_{\phi} <
7.69$. Since this violate a perturbative analysis of the wave
functions of ultra-cold neutrons, the upper bound on the coupling
constant $\beta$ should be substantially decreased. Using the
numerical values of the coefficients $C_k$ for $k = 6,7,8$ we get
\begin{eqnarray}\label{eq:19}
\hspace{-0.3in}\bar{C}_6 &=& 0.42 + 1.02\times
10^{-9}\,\beta\,a_{67}(n),\nonumber\\
\hspace{-0.3in}\bar{C}_7 &=& 0.66 + 0.20\times
10^{-9}\,\beta\,a_{67}(n),\nonumber\\
\hspace{-0.3in}\bar{C}_8 &=& 0.55 - 1.02\times
10^{-9}\,\beta\,a_{67}(n).
\end{eqnarray}
---------------new formulas for n=2 and h=47----------------
\begin{eqnarray}\label{eq:19}
\hspace{-0.3in}\bar{C}_5 &=& 0.15 + 0.37\times
10^{-9}\,\beta\nonumber\\
\hspace{-0.3in}\bar{C}_6 &=& 0.42 + 0.13\times
10^{-9}\,\beta\nonumber\\
\hspace{-0.3in}\bar{C}_7 &=& 0.66 - 1.80\times
10^{-9}\,\beta\nonumber\\
\hspace{-0.3in}\bar{C}_8 &=& 0.55 + 2.01\times
10^{-9}\,\beta\nonumber\\
\hspace{-0.3in}\bar{C}_{10} &=& 0.17 - 0.06\times
10^{-9}\,\beta\nonumber\\
\end{eqnarray}

A reasonable upper bound on the coupling constant $C_{\phi}$
compatible with a perturbative analysis of contributions of a
chameleon field is
\begin{eqnarray}\label{eq:20}
\hspace{-0.3in} C_{\phi} a_{67}(n) = 1.54\times
10^{-9}\,\beta\,a_{67}(n) \le 0.1.
\end{eqnarray}
This gives a new upper bound on the chameleon--matter coupling
constant $\beta \le 6.5 \times 10^8$, which is of order of magnitude
smaller compared with the upper bound $\beta < 5\times 10^9$, obtained
in \cite{Jenke2012} from the transition frequencies of quantum
gravitational states of ultra-cold neutron, confined between two
mirrors. For the upper bound Eq.(\ref{eq:20}) the coefficients
$\bar{C}_k$ with $k = 6,7,8$ are
\begin{eqnarray}\label{eq:21}
\hspace{-0.3in}0.42 \le &\bar{C}_6& \le 0.49,\nonumber\\
\hspace{-0.3in} 0.66 \le &\bar{C}_7&
\le 0.67,\nonumber\\
\hspace{-0.3in}0.55 \ge &\bar{C}_8& \ge 0.48
\end{eqnarray}
with the sum of the squared absolute values equal to $|\bar{C}_6|^2 +
|\bar{C}_7|^2 + |\bar{C}_8|^2 = 0.92$. 

The estimates of the coefficients $\bar{C}_k$ with $k = 6,7,8$ may be
extracted from the experimental data on a free fall of ultra-cold
neutrons in the spatial region $z \ge - D$ above a mirror with the
wave function
\begin{eqnarray}\label{eq:22}
|\Psi(z, x, t)|^2 &=& |\bar{C}_6\Psi_6(z,t) + \bar{C}_7\Psi_7(z,t) +
\bar{C}_8\Psi_8(z,t)|^2\nonumber\\ &&\times\,\frac{1}{\sqrt{\pi
    \Delta^2_t}}\,e^{\,\textstyle -\frac{(x - v_0t)^2}{2 \Delta^2_t}}.
\end{eqnarray}
The experimental data on a free fall of ultra-cold neutrons may be
fitted by only one parameter $\kappa = 0.20\times
10^{-9}\,\beta\,a_{67}(n)$, defining the coefficients $\bar{C}_k$ as
follows $\bar{C}_6 = 0.42 + 5\kappa$, $\bar{C}_7 = 0.66 + \kappa$ and
$\bar{C}_8 = 0.55 - 5\kappa$, respectively.



\section{Discussion}
\label{sec:discussion}

We have analysed a quantum free fall of ultra-cold neutrons from the
gap of a spatial region $\textstyle z^2 \le \frac{d^2}{4}$ between two
mirrors with $d = 25.5\,{\rm \mu m}$ into the spatial region above a
mirror, located at $ z = - D$ below the lower mirror at $D = 47\,{\rm
  \mu m}$. Between two mirrors ultra-cold neutrons are prepared in the
ground quantum gravitational state and move with a velocity $v_0 =
6\,{\rm m/s}$ in the $x$--direction with the wave function, taken in
the form of a Gaussian wave packet. For a depth $\textstyle D -
\frac{d}{2} = 47\,{\rm \mu m}$ of a spatial region of a quantum free
fall we have shown that the wave function of the $z$--degrees of
freedom of ultra-cold neutrons can be approximated by a superposition
of three excited quantum gravitational states with the principal
quantum numbers $k = 6, 7$ and $8$ with coefficients $C_6 = 0.42$,
$C_7 = 0.66$ and $C_8 = 0.55$, respectively. The probability of an
observation of ultra-cold neutrons in such a mixed quantum
gravitational state is $P = |C_6|^2 + |C_7|^2 + |C_8|^2 = 92$.  Using
perturbation theory we have calculated the first order corrections to
the wave functions of ultra-cold neutrons in the spatial region of a
quantum free fall. A numerical analysis shows that the main
contributions to the wave function of the $z$--degrees of freedom of
ultra-cold neutrons appear only to three excited quantum gravitational
with the principal quantum numbers $k = 6,7$ and $8$ and from these
excited states. The contributions of a chameleon field changes the
coefficients $C_k$ with $k = 6, 7$ and $8$ as follows $C_6 \to
\bar{C}_6 + 5\kappa$, $C_7 \to \bar{C}_7 \to 0.66 + \kappa$ and $C_8
\to \bar{C}_8 = 0.55 - 5\kappa$, where $\kappa = 0.20\times
10^{-9\,\beta\,a_{67}}(n)$ and $a_{67}/n)$ is a function of the power
$n$ of the potential of a self--interaction of a chameleon field $0.2
\ge a_{67} \ge 0.1$ for $n \in [2,10]$. A perturbative analysis of the
contributions of a chameleon field assumes that the effective coupling
constant $C_{\phi} = 1.54 \times 10^{-9}\,\beta$ is much smaller
compared with unity, i.e. $C_{\phi} \ll 1$. For the estimate $\beta <
5 \times 10^9$ we obtain $C_{\phi} < 7.7$. Since this value is not
compatible with a perturbative analysis of a chameleon field and
violates the unitarity condition on the coefficients $\bar{C}_6$,
$\bar{C}_7$ and $\bar{C}_8$, i.e. $|\bar{C}_k|^2 < 1$ and
$|\bar{C}_6|^2 + |\bar{C}_7|^2 + |\bar{C}_8|^2 < 1$, the estimate
$\beta < 5 \times 10^9$ should be improved in agreement with a
perturbative analysis of chameleon field and a unitarity condition. A
reasonable theoretical upper bound on the effective coupling constant
$C_{\phi} \le 0.1$ compatible with a unitarity condition of the
coefficients $\bar{C}_6$, $\bar{C}_7$ and $\bar{C}_8$ leads to a new
constraint on the chameleon--matter coupling constant $\beta$,
i.e. $\beta \le 6.5 \times 10^8$, which is of order of magnitude
smaller compared with the estimate $\beta < 5\times 10^9$
\cite{Jenke2012}. A new upper bound $\beta \le 6.5 \times 10^8$ is
compatible with the experimental data on the transition frequencies of
quantum gravitational states of ultra-cold neutrons \cite{Jenke2012}
at the level of a sensitivity $S = \Delta \omega/\omega \le
0.6\,\%$. For the theoretical transitions $|6\rangle \to |7\rangle$
and $|7\rangle \to |8\rangle$ the upper bound $\beta \le 6.5 \times
10^8$ may be observed for the sensitivities $S_{67} \le 0.6\,$ and
$S_{78} \le 1.5\,\%$, respectively. 

The stability of the new upper bound $\beta \le 6.5\times 10^{-8}$ one may prove by varying the depth of a quantum free fall region. Indeed, decreasing the depth of a quantum free fall region to $D = 30\,{\rm \mu m}$ leads to the following approximation of the wave function $\psi_1(z)$
owing approximation of the wave function $\psi_1(z,t)$
\begin{eqnarray}\label{eq:23}
\hspace{-0.3in}\psi_1(z,t) = C_3\Psi_3(z,t) + C_4\Psi_4(z,t) + C_5\Psi_5(z,t)
\end{eqnarray}
with the coefficients equal to $C_3 = 0.28$, $C_4 = C_5 = 0.66$. In
the region of a quantum free fall ultra-cold neutrons are in the mixed
state, described by the wave function Eq.(\ref{eq:23}), with a
probability $P = |C_3|^2 + |C_4|^2 + |C_5|^2 = 0.95$. The
contributions of a chameleon field are given by $\bar{C}_3 = 0.28 +
0.99\,\kappa$, $\bar{C}_4 = 0.66 - 1.22\,\kappa$ and $\bar{C}_4 = 0.66
+ 1.22\,\kappa$, where $\kappa = 1.54\times
10^{-9}\,\beta\,a_{54}(n)$. The function $a_{54}(n)$ varies over the
region $0.06 \le a_{54}\le 0.03$ for $n\in [2,10]$. One may show that
the upper bound $\beta < 5 \times 10^9$, obtained in \cite{Jenke2012},
violates a unitarity condition, whereas a new upper bound $\beta \le
6.5 \times 10^8$, obtained above, does not violate a unitarity. It is
also compatible with a perturbative analysis of the contributions of a
chameleon field. At $\beta \le 6.5 \times 10^8$ the coefficients
$\bar{C}_k$ are defined by the inequalities $0.28 \le \bar{C}_3 \le
0.33$, $0.66 \ge \bar{C}_4 \ge 0.59$ and $0.66 \le \bar{C}_3 \le 0.70$
and describe the mixed state Eq.(\ref{eq:23}) of ultra-cold neutrons
with the probability $P = |\bar{C}_3|^2 + |\bar{C}_4|^2 +
|\bar{C}_5|^2 = 0.95$.

An increase of the depth of the quantum free fall region, for example
to $D = 70\,{\rm \mu m}$ leads to a mixed quantum gravitational state
of ultra-cold neutrons, which should contain a complete set of pure
quantum gravitational states $\Psi_k(z,t)$. A truncated wave function,
containing the contributions of the state $\Psi_k(z,t)$ with the
largest coefficients, takes the form
\begin{eqnarray}\label{eq:24}
\hspace{-0.3in}&&\psi_1(z,t) = C_{10}\Psi_{10}(z,t) + C_{11}\Psi_4(z,t)
+ C_{13}\Psi_{13}(z,t),\nonumber\\
\hspace{-0.3in}&&
\end{eqnarray}
with $C_{10} = 0.35$, $C_{11} = 0.58$ and $C_{13} = - 0.14$,
respectively. The probability of an observation of ultra-cold neutrons
in the state with the wave function Eq.(\ref{eq:24}) is $P =
|C_{10}|^2 + |C_{11}|^2 + |C_{13}|^2 = 0.48$. For the analysis of a
contribution of a chameleon field we may neglect the contribution of
the wave function $\Psi_{13}(z,t)$. The probability of the state with
the wave function $\psi_1(z,t) = C_{10}\Psi_{10}(z,t) +
C_{11}\Psi_4(z,t)$ is $P = |C_{10}|^2 + |C_{11}|^2 = 0.46$. The
contribution of a chameleon field changes the coefficients $C_{10}$
and $C_{11}$ as follows $\bar{C}_{10} = C_{10} + C_{11}\,\kappa$ and
$\bar{C}_{11} = C_{11} - C_{10}\,\kappa$, where $\kappa = 1.54\times
10^{-9}\,\beta\,a_{10,11}(n)$ with $0.28 \ge a_{10,11}(n) \ge 0.10$
for $n \in [2,10]$. For $\beta \le 6.5 \times 10^8$ we get $0.35 \le
\bar{C}_{10} \le 0.51$ and $0.58 \ge \bar{C}_{11} \ge 0.48$.

The proposed analysis of the dependence of the upper bound of the
chameleon--matter coupling constant $\beta$ on the depth of a quantum
free fall region of ultra-cold neutrons confirms a stability of the
estimate $\beta \le 6.5 \times 10^8$. Of course, such a theoretical
constraint should be improved
experimentally by fitting the experimental data on the QBB experiments
by only one parameter $\kappa$.

\section{Conclusion}
\label{sec:conclusion}

This study refines the upper bound on the chameleon–matter coupling constant ($\beta\le 6.5 \times 10^8$), advancing our understanding of scalar field theories and their role in cosmic acceleration. By leveraging the unique properties of ultra-cold neutrons in qBounce experiments, we demonstrate the capability of laboratory-based setups to probe fundamental physics with unprecedented precision.

The findings highlight the utility of ultra-cold neutron experiments as precise laboratories for testing fundamental theories in physics. By providing a significantly tighter upper bound on the chameleon–matter coupling constant, this research underscores the sensitivity of quantum gravitational systems to subtle scalar field interactions.

These results have profound implications for the role of chameleon fields as drivers of cosmic acceleration. The refined constraints on $\beta$ challenge the viability of certain parameter ranges within chameleon field theories, narrowing the scope of models compatible with observed cosmic acceleration. By demonstrating the sensitivity of terrestrial experiments to such scalar field interactions, this work strengthens the case for chameleon fields as potential contributors to dark energy while also placing stringent limits on their behaviour.

One critical implication is the potential of qBounce setups to distinguish between competing dark energy models. The refined constraints can aid in differentiating chameleon field effects from alternative mechanisms driving cosmic acceleration. Moreover, the methodology described herein can be adapted to investigate other scalar fields, broadening its applicability in particle physics and cosmology.

Future work should focus on enhancing experimental sensitivities, particularly in detecting higher-order perturbations induced by chameleon fields. Exploring variations in experimental parameters, such as mirror separation or neutron velocity, may yield deeper insights into scalar field dynamics. The integration of complementary observational data from astrophysical phenomena could further validate the theoretical models tested here.

The implications extend beyond chameleon fields, offering insights into scalar field dynamics, quantum mechanics, and gravitational phenomena. These findings lay the groundwork for future experimental and theoretical advancements, highlighting the interplay between cosmology and quantum physics in addressing some of the Universe's most profound questions.

\begin{acknowledgments}
We want to thank our dear colleague  Andrey Nikolaevich Ivanov, who was the main investigator of this work until he sadly passed away on December 18, 2021. We see it as our professional and personal duty to honor his legacy by continuing to publish our collaborative work. Andrey was born on June 3, 1945 in what was then Leningrad. Since 1993 he was a university professor at the Faculty of Physics, named "Peter The Great St. Petersburg Polytechnic University" after Peter the Great. Since 1995 he has been a guest professor at the Institute for Nuclear Physics at the Vienna University of Technology for several years and has been closely associated with the institute ever since. This is also were we met Andrey and have been collaborating with him closely over more than 20 years resulting in 40 scientific publications, see also~\cite{Ivanov:2013fca, Ivanov:2018qen, Ivanov:2018ngi, Ivanov:2018olo, Ivanov:2018yir, Ivanov:2019bqr, Ivanov:2020ybx, Ivanov:2021bae, Ivanov:2021lji, Ivanov:2021yhl, Ivanov:2021xkm, Altarawneh:2024psz}. 
We will miss Andrey as a personal friend and his immense wealth of ideas, scientific skills and his creativity. See also the \href{https://www.tuwien.at/en/phy/ati/news/test}{official obituary} for Andrey Nikolaevich Ivanov.
\end{acknowledgments}

\subsection*{Appendix A: Wave functions of ultra-cold neutrons in the 
gravitational field of the Earth between two mirrors and above a
mirror} \renewcommand{\theequation}{A-\arabic{equation}}
\setcounter{equation}{0}\label{app:A}

The wave function $\psi_1(z)$ of the ground quantum gravitational
state of ultra-cold neutrons between two mirrors in the spatial region
$\textstyle z^2 \le \frac{d^2}{4}$ is equal to \cite{Ivanov2012}
\begin{eqnarray}\label{labelA.1}
\hspace{-0.3in}\psi_1(z) &=& \Big\{{\rm
  Ai}(\xi_k){\rm Bi}\Big[\frac{1}{\ell_0
}\Big(\frac{d}{2} + z\Big) + \xi_1\Big]\nonumber\\
\hspace{-0.3in} &-& {\rm Ai}\Big[\frac{1}{\ell_0}
\Big(\frac{d}{2} + z\Big) + \xi_1\Big]{\rm Bi}(\xi_1)\Big\}\nonumber\\
\hspace{-0.3in}&&\times \Big\{\int^{+d/2}_{-d/2} \Big|{\rm
  Ai}(\xi_1){\rm Bi}\Big[\frac{1}{\ell_0 }\Big(\frac{d}{2} + z\Big)+
  \xi_1\Big]\nonumber\\
\hspace{-0.3in}&-& {\rm
    Ai}\Big[\frac{1}{\ell_0 }\Big(\frac{d}{2} + z\Big) +
  \xi_1\Big]{\rm Bi}(\xi_1) \Big|^2 dz\Big\}^{-1/2}.
\end{eqnarray}
The binding energy $E_1 = - m g \ell_0 \xi_1 = 1.41\,{\rm peV}$ of the
ground quantum gravitational state is given by the equation
\begin{eqnarray}\label{labelA.2}
\hspace{-0.3in}{\rm Ai}(\xi_1){\rm Bi}\Big(\frac{d}{\ell_0} +
\xi_1\Big) - {\rm Ai}\Big(\frac{d}{\ell_0} + \xi_1\Big){\rm Bi}(\xi_1)
= 0
\end{eqnarray}
with the root $\xi_1 = - 2.34497$, calculated at $d = 25.5\,{\rm \mu
  m}$, $\ell_0 = (2 m^2 g)^{-1/3} = 5.87\,{\rm \mu m}$, $m =
939.5654\,{\rm MeV}$ and $g = 980.6\,{\rm cm/s^2}$ \cite{PDG12}.

Above a mirror in the spatial region $z\ge - D$ the wave functions of
ultra-cold neutrons are
\begin{eqnarray}\label{labelA.3}
\hspace{-0.3in}\Psi_k(z) = \frac{\displaystyle {\rm Ai}\Big(\frac{D +
      z}{\ell_0} + \zeta_k\Big)}{\displaystyle
  \sqrt{\int^{+\infty}_{-D} \Big|{\rm Ai}\Big(\frac{D + z}{\ell_0 } +
      \zeta_k\Big)\Big|^2 dz}},
\end{eqnarray}
where $\zeta_k$ are the roots of the equation ${\rm Ai}(\zeta_k) = 0$
\cite{qBouncer}. The binding energies of the first 15 quantum
gravitational states of ultra-cold neutrons and the coefficients $C_k$
for $k =1, 2,\ldots, 15$ are adduced in Table I.



\begin{table*}
\begin{tabular}{|c|c|c|c|c|c|c|c|}
\hline k&  $\zeta_k$& $E_k$ & $C_k (h=31\mu m)$& $C_k (h=39\mu m)$& $C_k (h=47\mu m)$& $C_k (h=55\mu m)$& $C_k (h=63\mu m)$\\\hline 
1&-2.33811 &1.407 &$2\times10^{-3}$&$1.08\times10^{-4}$& $+4.03\times10^{-6}$&$1.05\times10^{-7}$&$1.99\times10^{-9}$\\\hline 
2&-4.08795 & 2.461 &$3.79\times10^{-2}$&$3.66\times10^{-3}$& $+2.23\times 10^{-4}$&$9.27\times10^{-6}$&$2.67\times10^{-7}$\\\hline
3&-5.52056 & 3.324 &${\bf0.23}$&$0.04$& $+3.89\times 10^{-3}$&$2.44\times10^{-4}$&$1.03\times10^{-5}$\\\hline 
4&-6.78671 & 4.086 &${\bf0.61}$&${\bf0.20}$&$+3.22\times 10^{-2}$&$3.08\times10^{-3}$&$1.8\times10^{-4}$\\\hline 
5&-7.94413 & 4.782 &${\bf0.70}$&${\bf0.52}$&${\bf+0.15}$&$0.02$&$1.9\times10^{-3}$\\\hline
6&-9.02265 & 5.432 &${\bf0.17}$&${\bf0.71}$&${\bf+0.42}$&$0.10$&$0.01$\\\hline
7&-10.04020 & 6.044 &${\bf-0.18}$&${\bf0.36}$&${\bf+0.66}$&${\bf0.30}$&$0.06$\\\hline
8&-11.00850 & 6.627 &$0.03$&${\bf-0.13}$&${\bf+0.55}$&${\bf0.58}$&${\bf0.20}$\\\hline
9&-11.93600 & 7.186 &$0.02$&$-0.08$&$+0.06$&${\bf0.65}$&${\bf0.44}$\\\hline
10&-12.82880 & 7.723&$-0.04$&$0.08$&${\bf-0.17}$&${\bf0.31}$&${\bf0.64}$\\\hline
11&-13.69150 & 8.242&$0.05$&$-0.04$&$+0.02$&${\bf-0.11}$&${\bf0.53}$\\\hline
12&-16.13270 & 9.712&$0.02$&$-0.04$&$+0.03$&$4.69\times10^{-3}$&$-0.05$\\\hline
13&-16.90560 & 10.177&$-0.03$&$0.03$&$+0.01$&$-0.04$&${\bf0.07}$\\\hline
14&-19.12640 & 11.514&$0.01$&$0.03$&$-0.01$&$-0.03$&$0.05$\\\hline
15&-19.83810 & 11.943&$-1.38\times10^{-3}$&$-0.02$&$-0.01$&$0.04$&$-0.02$\\\hline
\multicolumn{3}{|c|}{$P_{total}=\sum_k|C_k|^2$}&98.67\%&98.68\%&98.01\%&96.80\%&94.52\%\\\hline
\multicolumn{3}{|c|}{$P_{considered}=\sum_i|{\it\bf C_i}|^2$}&97.8\%&96.72\%&97.46\%&95.29\%&93.62\%\\\hline
\end{tabular}
\caption{The binding energies of the first 15 quantum gravitational
  states of ultra-cold neutrons in the gravitational field of the
  Earth above a mirror and the coefficients $C_k$ for $k =1,2,\ldots,
  15$ and different heights $h=31\mu m,39\mu m,47\mu m,55\mu m,63\mu m$. The last two rows show the probabilities $P_{total}=\sum_k|C_k|$ for $k\in[1,2,\ldots,15]$ and $P_{considered}$ for the $C_k$s plotted in bold, which are taken into account for the calculation of transition coefficients $a_{k'k}(n)$.}
\end{table*}

\end{document}